\documentclass[letterpaper,10pt,conference]{ieeeconf} 
\IEEEoverridecommandlockouts 
\makeatletter{\normalsize } 

\let\AND\@undefined
\makeatother

\makeatletter
\let\NAT@parse\undefined
\makeatother
\usepackage[utf8]{inputenc}
\newcommand{\matlab}{\textsc{matlab}}
\usepackage{epsfig} 
\usepackage{mathptmx} 
\usepackage{amsmath} 
\usepackage{amssymb}
\usepackage{graphicx}
\usepackage{psfrag, units, booktabs}	
\usepackage{pdfpages}
\usepackage{float}	
\usepackage{algorithmic}
\usepackage{algorithm}
\usepackage{amsmath}
\usepackage{mathtools}
\usepackage{pgfplots}
\usepackage{eurosym, color, subfloat, xfrac}
\usepackage{placeins}
\usepackage{tikz}
\usetikzlibrary{shapes,arrows}
\usepackage{circuitikz}
\usepackage{bytefield}
\usetikzlibrary{decorations.pathreplacing}
\usepackage{adjustbox}
\usepackage[binary-units=true]{siunitx}
\usepackage{multirow}
\usepackage[colorinlistoftodos]{todonotes}
\usepackage{fancyvrb}
\usepackage{xspace}
\usepackage{pgfplots}
\usepackage{eurosym, color, subfloat, xfrac}
\usepackage{placeins}
\usepackage{tikz}
\usetikzlibrary{shapes,arrows}
\usetikzlibrary{decorations.pathreplacing}
\usepackage{relsize}
\usepackage{listings}
\usepackage{soul}
\usepackage{hyperref}
\hypersetup{
colorlinks=true,
linkcolor=blue,
citecolor = blue,
filecolor=magenta, 
urlcolor=black,
}
\usepackage{cite} 

\newcommand\cpp{C\nolinebreak[4]\hspace{-.05em}\raisebox{.4ex}{\relsize{-3}{\textbf{++~}}}}

\usepackage{expl3}
\ExplSyntaxOn
\newcommand\latinabbrev[1]{
\peek_meaning:NTF . {
#1\@}%
{ \peek_catcode:NTF a {
	#1.\@ }%
{#1.\@}}}
\ExplSyntaxOff

\def\ie{\latinabbrev{i.e}}

\title{\LARGE \bf Embedded Model Predictive Control Using Robust Penalty Method}
\author{Abhijith Sharma$^{1}$, Chaitanya Jugade$^{1}$, 
Shreya Yawalkar$^{1}$, Vaishali Patne$^{1}$, \\ Deepak Ingole$^{2}$, and Dayaram Sonawane$^{1}$ 
\thanks{$^{1}$ College of Engineering Pune, Shivajinagar $411005$, India. {\tt\small \{\href{mailto:abhijiths16.instru@coep.ac.in}{abhijiths16.instru}, \href{jugadeck18.instru@coep.ac.in}{jugadeck18.instru}, \href{yawalkarsu19.instru@coep.ac.in}{yawalkarsu19.instru}, \href{pva18.instru@coep.ac.in}{pva18.instru},
\href{dns.instru@coep.ac.in}{dns.instru\}}@coep.ac.in}} 
\thanks{$^{2}$ KU Leuven, Department of Mechanical Engineering, Leuven, Belgium. {\tt\small \href{mailto:deepak.ingole@kuleuven.be}{deepak.ingole@kuleuven.be}}}	
}
\begin{document}
\maketitle
\thispagestyle{empty}
\pagestyle{empty}
\begin{abstract}	
\label{abstract}
Model predictive control (MPC) has become a hot cake technology for various applications due to its ability to handle multi-input multi-output systems with physical constraints. The optimization solvers require considerable time, limiting their embedded implementation for real-time control. To overcome the bottleneck of traditional quadratic programming (QP) solvers, this paper proposes a robust penalty method (RPM) to solve an optimization problem in a linear MPC. The main idea of RPM is to solve an unconstrained QP problem using Broyden–Fletcher–Goldfarb–Shannon (BFGS) algorithm. The beauty of this method is that it can find optimal solutions even if initial conditions are in an infeasible region, which makes it robust. Moreover, the RPM is computationally inexpensive as compared to the traditional QP solvers. The proposed RPM is implemented on resource-limited embedded hardware (STM32 microcontroller), and its performance is validated with a case study of a citation aircraft control problem. We show the hardware-in-the-loop co-simulation results of the proposed RPM and compared them with the active set method (ASM) and interior point method (IPM) QP solvers. The performance of MPC with the aforementioned solvers is compared by considering the optimality, time complexity, and ease of hardware implementation. Presented results show that the proposed RPM gives the same optimality as ASM and IPM, and outperforms them in terms of speed. 
\end{abstract}
\section{INTRODUCTION}	
\label{sec:intro}
The model predictive control (MPC) has played a comprehensive role in process industries. ~\cite{garcia:1989:model}.
This algorithm has been widely accepted in process industries. The underlying reason for this popularity is the less human intervention requirement and ease in tuning for operators. 
It has been implemented under restricted conditions a myriad of times~\cite{mayne:2014:model,badgwell:2013:MPC} in petrochemical, oil, and gas, and food processing industries~\cite{mayne:2014:model}. 
This is possible because of its capability of handling multi-input and multi-output (MIMO) systems with constraints on inputs, states, and outputs~\cite{borrelli:2017:predictive}. 
It has been successfully imbibed by aerospace, autonomous vehicles, automobile sectors in the present scenario~\cite{Badgwell2013}. 

MPC is an optimal control algorithm wherein the control action is obtained by solving a constrained convex optimization problem (generally convex quadratic programming (QP) problem) over the finite-time interval for the system's current state at each sampling time. In general, MPC formulation leads to three components of the system's response: heading of process dynamics (state estimation), final destination (steady-state target optimization), and the best set of control (input) that would drive the states to target. The initial control (input) of the sequence obtained is implemented, and then the entire calculation is repeated for the subsequent control cycles~\cite[Chapter 12]{borrelli:2017:predictive}. 
This control technique of repeatedly solving a constrained problem over moving time to choose a control horizon is called receding horizon control (RHC). This introduces inherent negative feedback to the system, allowing automatic compensation of disturbances in the systems~\cite[Chapter 12]{borrelli:2017:predictive}.

There are several state-of-the-art solvers developed to solve the underlying QP problem in an MPC. Examples of such solvers are CVXGEN, $\mu$AO-MPC, ECOS, qpOASES, etc. CVXGEN generates custom c code for embedded applications, but cannot explicitly handle feasibility and constraints.~\cite{mattingley:2012:cvxgen}. The $\mu$AO-MPC offers low memory utilization for real-time linear MPC implementation, but the execution time is slow. ECOS is based on a primal-dual IPM.
The majority of the solvers employ algorithms that are based on the active set method (ASM) and the interior point method (IPM)~\cite{mattingley:2012:cvxgen, wang:2009:fast, domahidi:2013:ecos, chu:2013:code, ferreau:2014:qpoases}. The active set method considers the set of active constraints for each iteration to solve the QP problem. As a result, computation time is directly proportional to the number of active constraints. In contrast, the computation time of IPM is relatively constant, regardless of the number of active constraints. This time can be large enough compared to ASM in cases of a small QP problem with fewer constraints and variables~\cite{bartlett:2000:active}. It has been shown that embedded implementation of the IPM and ASM is limited to the small problems and needs more resources, i.e., memory, clock cycles, etc. Authors in~\cite{lau:2009:comparison} presented the approaches of solving QP problems using ASM and IPM methods and compared the performances for an FPGA implementation for MPC applications. They have mentioned that ASM can make the system unstable due to numerical errors for the control system applications compared to IPM. To mitigate this problem many authors worked on different approaches. Authors in~\cite{jugade:2020:framework} presented the novel idea to use the posit number system to reduce numerical error in ASM and improving the performance. In~\cite{arnstrom:2021:unifying}, the authors proposed an interesting method to identify the worst-case number of iterations required for the primal and dual active-set algorithms to reach optimality. Authors in~\cite{ding:2016:embedded} introduced the convergence depth control method into the interior-point method to accelerate the QP solving process for embedded MPC implementation.

For large scale problems, alternating method of multipliers (ADMM) is preferred. However, due to large data, it can become complex for implementation~\cite{sutor:2016:admm}. Another method that is used for optimization of QP problem is gradient descent method. These methods use sensitivity information to evaluate the search directions~\cite{kogel:2011:gradient}. 
The real-time dynamic systems introduce the principle challenge for the implementation of MPC. These systems demand a high sampling rate with embedded implementation on hardware with limited resources. To overcome the bottleneck of traditional QP solvers, we propose a robust penalty method (RPM)-based linear MPC and its microcontroller implementation. Because of its simplicity and intuitive appeal, this approach is often used in practice. The idea of the penalty method is to convert the original constrained QP problem to an unconstrained optimization problem. This reduces the solving time for the problem. We have developed a robust penalty method by incorporating Broyden–Fletcher–Goldfarb–Shannon (BFGS) algorithm to solve the unconstrained QP problem. The concept of the penalty method~\cite[Chapter 17]{nocedal:2006:numerical} is not new but has some limitations. The original penalty method stops when it reaches the boundary of constraints. This happens because no constraints will be violated when it is inside a feasible region; hence no penalty function is used. But, there can be a scenario when the optimal value is completely inside the feasible region and not at the boundary. To handle this problem, we propose to solve the unconstrained QP problem without penalty. 
This will give us an optimal solution even if the initial guess is in the infeasible region.
The proposed penalty method is implemented on an STM32 microcontroller. 
The hardware-in-the-loop (HIL) co-simulation results of the proposed penalty method are presented for a benchmark QP problem and a case study of citation aircraft control problem using MPC. The main contribution of this paper is to develop a robust penalty method and its embedded implementation to highlight the performance of the penalty approach in solving the MPC problem as well as in standalone QP problems.

The paper is organized as follows: Section~\ref{sec:lmpc} describes the MPC problem formulation. In Section~\ref{sec:epm}, a detailed description of the robust penalty method is presented. In Section~\ref{sec:embedded:lmpc}, embedded implementation of RPM, ASM, and IPM for linear MPC is discussed. Section~\ref{sec:result} presents the hardware-in-the-loop (HIL) results of all QP methods for the benchmark QP problem and citation aircraft control problem. At the end, conclusion with a summary of the work and possible future scope is stated in Section~\ref{sec:conclusion}.
\section{Linear Model Predictive Control}
\label{sec:lmpc}
This section describes the linear model predictive controller and its QP problem formulation for reference tracking. 
\subsection{Prediction Model}
\label{subsec:mpc:model}
Consider a discrete-time version of the linear-time invariant (LTI) system~\eqref{eq:dlti},
\begin{subequations}
	\label{eq:dlti}
	\begin{align}
		\label{eq:dlti:state}
		x(t+T_s) &= Ax(t)+Bu(t),\\
		\label{eq:dlti:output}
		y(t)&= Cx(t) + Du(t),
	\end{align}
\end{subequations}
where $x(t)\in \mathbb{R}^{n}$ is the system state vector, $u(t)\in \mathbb{R}^{l}$ is the system input vector, $y(t)\in \mathbb{R}^{m}$ is the system output vector, and $T_s$ is the sampling time. Moreover, $A \in\mathbb{R}^{n\times n}$, $B\in\mathbb{R}^{n\times l}$, $C\in\mathbb{R}^{m\times n}$, and $D \in \mathbb{R}^{m \times l}$ are system matrices. We assume that the pair $(A, B)$ is stabilizable, and $(C, A)$ is detectable.
\subsection{Optimal Control Problem}
\label{subsec:mpc:ocp}
A constrained finite-time optimal control (CFTOC) problem considering model in~\eqref{eq:dlti} for reference tracking can be represented as follows:
\begin{subequations}
\label{eq:cftoc}
\begin{align}
\label{eq:cftoc:obj}
\hspace{-8mm}\min_{U} \sum ^{N-1}_{k=0} (y_{k}-y_{\text{r},k})^{T}Q(y_{k}-y_{\text{r},k}) + \Delta u^{T}_{k}R\Delta u_{k},
\end{align}
\vspace{-5mm}
\begin{align}\nonumber
& \text{s.t.} \\
\label{eq:cftoc:state}
\vspace{-5mm}
&x_{k+T_s} =Ax_{k} +Bu_{k} ,&k=0,\dots,N-1,\\
\label{eq:cftoc:output}
&y_{k} =Cx_{k}+ Du_k,&k=0,\dots,N-1,\\
\label{eq:cftoc:deltau}
&\Delta u_k = u_k - u_{k-1}, &k=0,\dots,N-1,\\
\label{eq:cftoc:const:x}
   &x_{k} \in {\Large \textit{x}}, &k=0,\dots,N-1,\\
\label{eq:cftoc:const:u}
	&u_{k} \in {\Large \textit{u}} ,&k=0,\dots,N-1,\\
\label{eq:cftoc:const:y}
	&y_{k} \in {\Large \textit{y}} ,&k=0,\dots,N-1,\\
\label{eq:cftoc:ini:u}
&u_{-1} = u(t-T_s), \\
\label{eq:cftoc:ini:x}
& x_{0} = x(t),
\end{align}
\end{subequations}
where $Q \in \mathbb{R}^{n \times n}$ and $R \in \mathbb{R}^{l \times l}$ are the weighting matrices, with $Q \succeq 0$ and $R \succ 0$ to be positive definite. We denote by $N$ the prediction horizon, $x_{k+1}$ as the vector of predicted states at time instant $k$, $U = \{u_0, \ldots, u_{N-1}\}$ as the sequence of control actions, $y_{r,k}$ is the output reference trajectory to be tracked, and $x_0$ and $u_{-1}$ are the given initial conditions. States, inputs, and outputs belong to polytopic constraints $x \in {\Large \textit{x}} \subseteq \mathbb{R}^{n}$, $u\in {\Large \textit{u}} \subseteq \mathbb{R}^{l}$, and $y\in {\Large \textit{y}} \subseteq \mathbb{R}^{m}$ where ${\Large \textit{x}}$, ${\Large \textit{u}}$, and ${\Large \textit{y}}$ are polyhedral sets. The optimization is performed with respect to $U = \{u_0^\star,\dots,u_{N-1}^\star\}$. As per the concept of a receding horizon control (RHC), only the first optimized input, \ie, $u_0^\star$ is applied to the system in~\eqref{eq:dlti}, and the whole procedure is repeated at a subsequent time instant for a new value of the initial condition obtained using~\eqref{eq:cftoc:ini:u} and~\eqref{eq:cftoc:ini:x}.
\subsection{QP Problem Formulation}
\label{subsec:Qp:solvers}
The open-loop CFTOC problem in~\eqref{eq:cftoc} can be formulated as a general convex QP problem as given below. For the detailed formulation of the QP problem see~\cite{maciejowski:2002:predictive}. 
\begin{subequations}
	\label{eq:qp}
	\begin{align}
		\label{eq:qp:cost}
		\underset{U}{\text{min}} & \; \frac{1}{2} U^THU + f^TU, \\
		\label{eq:qp:const}
		\text{s.t.} \; 	& \; G U \le w,
	\end{align}
\end{subequations}
where $H \in \mathbb{R}^{lN \times lN}$ is the Hessian matrix, $f\in \mathbb{R}^{n \times lN},~G \in \mathbb{R}^{q\times lN},~\rm{and}~w \in \mathbb{R}^{q},$ are the matrices/vectors with $q$ as number of inequalities. The most general approach to solve QP problems is to use the active-set methods~\cite{fletcher:2013:practical}, interior-point methods~\cite{nocedal:2006:numerical}, and gradient-based~\cite{snyman:2005:practical} methods, which have shown good convergence and stability properties. However, we propose a penalty method which, in addition to having good convergence and stability will also perform well for the problem with a large number of constraints, unlike ASM. Moreover, the dependency of initial guess does not influence the convergence, which is a common issue with IPM. 
\section{Penalty Method} 
\label{sec:epm}
In this section, we describe the proposed robust penalty method-based QP solver aiming to develop fast embedded MPC. Consider the following QP problem:
\begin{subequations}
	\label{eq:augobj}
	\begin{align}
		\label{eq:augobj1}
		\underset{U}{\text{min}} & \; F = J + K^T P,\\
		\label{eq:augobj2}
		& \; P = (\max [ z , g ] )^2,\\
		\label{eq:augobj3}
		\text{where} : &g = GU - w,\\
		\text{ } \;	& \; J = \frac{1}{2} U^THU + f^TU,
	\end{align}
\end{subequations}
where $K \in \mathbb{R}^{q}$ is penalty gain vector with each element being the gain for elements in $P$. $P \in \mathbb{R}^{q}$ is the penalty function satisfying the following conditions: (i) $P$ is continuous, (ii) $P \geq 0$ for all $x \in E$, where $E$ is feasible set. The $z \in \mathbb{R}^{q}$ such that $z = [0, \dots, 0]$. The frequently used penalty function is as shown in~\eqref{eq:augobj2}. Hence, the objective function $F(U)$ will be an augmentation of original function $J$ and penalty term $P$.

It is evident that as $K$ is large, the corresponding $P$ has to be small when the objective function is minimized~\cite[chapter 17]{nocedal:2006:numerical}. A subsequent increase in $K$ at each iteration would bring a corresponding point in the feasible region minimizing objective function $F$. 
The idea is to penalize the minimization function every time a constraint is violated. The general practice is to start $K$ with $0.1$ and increase it $10$ times for the next iteration. Soft handling of constraints makes the computations fast. For safety-critical applications, tolerance can be adjusted to get the desired performance. if the solution lies on the boundary for some problems, the tolerance can be defined to set the acceptable limit of the error in the optimal solution. Thus, defining this tolerance will prevent the solution from oscillating between feasible and infeasible regions and never settling. 
The algorithm used for the RPM to solve the QP problem~\cite{luenberger:1984:linear} in a linear MPC is presented in~\ref{algo:rpm}.
	
\begin{algorithm}[htbp]
\caption{Robust penalty method using BFGS unconstrained QP solver.}
\label{algo:rpm}
\begin{algorithmic}
	\STATE {\bfseries Input:} $J, g, K, \delta$.
	\STATE {\bfseries Output:} $U_{N}, F_{\rm val}$.
	\STATE Choose arbitrary $U$ and slack tolerance = $\xi$.
	\STATE iteration = 0
	\REPEAT
	\IF{($g>\xi$)}
	\STATE $K = \{0.1,\dots,0.1\}$
	\ENDIF
	\STATE Update $K = K\times10^{(\rm iteration)}$
	\STATE [$U_N, F_{\rm val}$] = \textbf{BFGSSolve}($J, g, K, U, \delta$)\COMMENT {See Algorithm~\ref{algo:bfgs}}
	\IF{($g\leq\xi)$}	
	\IF{(Convergence tolerence $< \delta$)}
	\STATE $U_{N}$ is the optimum point in feasible region
	\ELSE
	\STATE $U_{N}$ is in feasible region but not the optimum point
	\ENDIF
	\ELSE
	\STATE $U_{N}$ not in feasible region
	\ENDIF
	\STATE $U$ = $U_{N}$
	\STATE iteration = iteration + 1
	\UNTIL{$U_{N}$ is not optimum}
\end{algorithmic}
\end{algorithm}
%

To solve the unconstrained QP problem in~\eqref{eq:augobj}, a Quasi-Newton BFGS algorithm similar to~\cite[chapter 10]{luenberger:1984:linear} is used with a few changes. Using this algorithm, we can maintain the positive definiteness of the Hessian matrix. This is done by defining $\tilde{H_k}$, which is a positive definite identity matrix. As the point moves to the optimal solution, the approximation converges to the original Hessian. The approximate Hessian is revised based on the gradient information of previous and current iteration. This method helps to avoid unfavorable results caused due to ill-conditioning of Hessian matrix when penalty $K$ increases.

Let objective function $F(U)$ have a quadratic model, as shown in~\eqref{eq:qf1}:
\begin{subequations}
	\label{eq:qf}
	\begin{align}
		\label{eq:qf1}
		&M= F(U_{k}) + (\triangledown F(U_{k}))^T p_k + {p_k}^T \tilde{H_k} p_k,\\
		\label{eq:qf2}
		&p_k = {\tilde{-H_k^{-1}}}\triangledown F(U_{k}).
	\end{align}
\end{subequations}
The $p_k$ that would guarantee the minimization of the convex quadratic model is given in~\eqref{eq:qf2}. 
Algorithm~\ref{algo:bfgs} presents the BFGS method used to solve unconstrained QP problems using Algorithm~\ref{algo:rpm}. 
\begin{algorithm}[htbp]
	\caption{BFGS QP solver (\textbf{BFGSSolve}) used in RPM to solve unconstrained QP}
	\label{algo:bfgs}
	\begin{algorithmic}
		\STATE {\bfseries Input:} $J, g, K$.
		\STATE {\bfseries Output:} $U_{N}, F_{\rm val}$.
		\STATE Initialize $\tilde{H_k}$ to identity matrix $\in\mathbb{R}^{n\times n}$, convergence tolerance $\delta$, and solution tolerance $\epsilon$.
		\WHILE {$(\epsilon> \delta)$}
		\STATE Define $b = -\triangledown(F(U))$
		\STATE \hspace{10mm} $g =\tilde{{H_k}} b$
		\STATE Obtain $t_k$ by backtracking algorithm
		\STATE Set $U_{k+1} = U_{k} + t_k p_k$

		\IF{(in infeasible region)}
		\STATE check $\epsilon =\mid \frac{(U_k)-(U_{K+1})}{U_k}\mid$
		\IF {($\epsilon < \alpha$)}
		\STATE break
		\ELSE
		\STATE Update $\tilde{H_k}$ 
		\STATE $k=k+1$
		\STATE Update $K$
		\STATE $U_k = U_{k+1}$
		\ENDIF
		\ELSE
		\STATE check $\epsilon = \mid \frac{(U_k)-(U_{K+1})}{U_k}\mid$
		\IF {($\epsilon < \alpha$)}
		\STATE $U_N = U_{k+1}$
		\ELSE
		\STATE $U_k = U_{k+1}$
		\STATE value of $K$ does not change
		\STATE $k=k+1$
		\ENDIF
		\ENDIF
		\ENDWHILE
	\end{algorithmic}
\end{algorithm}
The Hessian ($H$) matrix derived in MPC formulation~\eqref{eq:qp:cost} turns out to be symmetric positive definite. Due to the nature of $H$, the inverse of the $H$ matrix can be obtained through Cholesky factorization~\cite{higham:2009:cholesky} and forward/backward substitution.
As the final goal is to have a fast solver, the inexact line search is implemented here. 

%
%
%
%

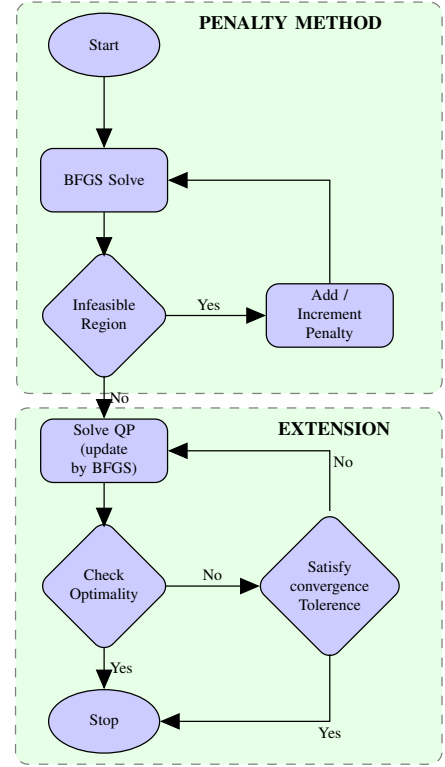
\begin{figure}[htb!]
\scalebox{1.2}{	\centering
\vspace*{-4mm}
\pgfdeclarelayer{background}
\pgfdeclarelayer{foreground}
\pgfsetlayers{background,main,foreground}
\tikzstyle{block} = [rectangle, draw, fill=blue!20, text centered, rounded corners, minimum height=4em,minimum width=8em,text width=5em]
\tikzstyle{bck} = [rectangle, draw, fill=blue!20, text centered, rounded corners, minimum height=4em,minimum width=8em,text width=6em]
\tikzstyle{bck1} = [rectangle, draw, fill=blue!20, text centered, rounded corners, minimum height=4em,minimum width=5em,text width=4em]
\tikzstyle{dia} = [diamond, draw, fill=blue!20, text centered, rounded corners, minimum height=2em,minimum width=4.5em,text width=4em]
\tikzstyle{diamon} = [diamond, draw, fill=blue!20, text centered, rounded corners, minimum height=3em,minimum width=6em,text width=5em]
\tikzstyle{ov} = [ellipse, draw, fill=blue!20, text centered, rounded corners, minimum height=4em,minimum width=7em,text width=4em]	
\tikzstyle{line} = [draw, -latex']
\begin{tikzpicture}[node distance = 3cm, auto,scale=0.5, every node/.style={transform shape}]
\node [ov,align=center,node distance = 1cm] (start) {Start};
\node [bck,below of=start, align=center] (solve) {BFGS Solve};
\node [diamon, below of=solve,align=center] (FR) {Infeasible Region};
\node [bck, below of=FR,align=center] (BFGS2) {Solve QP (update by BFGS)};	
\node [bck, right of=FR,align=center,node distance = 5cm] (pen) {Add / Increment Penalty};
\node [diamon, below of=BFGS2,align=center] (IFR) {Check Optimality};
\node [diamon, right of=IFR,align=center,node distance = 5.0cm] (ST) {Satisfy convergence Tolerence};
\node [ov,below of=IFR, align=center] (stop) {Stop};
\draw [-triangle 45] (start) -- (solve);
\draw [-triangle 45] (solve) -- ++(0,-1.7);  
\draw [-triangle 45] (FR.south)++(0,0.15) -- node{No}(BFGS2);
\draw [-triangle 45] (BFGS2) --  ++(0,-1.7);  
\draw [-triangle 45] (IFR.center)++(0,-1.35) --node{Yes} (stop);	
\draw [-triangle 45] (FR.center)++(1.35,0) -- node{\hspace{-3mm}Yes}(pen);
\draw [-triangle 45] (IFR.center)++(1.35,0) --node{No}++(2.1,0);    
\draw [-triangle 45] (pen) |- (solve);
\draw [-triangle 45] (ST) |-node{\hspace{6mm}No} (BFGS2);
\draw [-triangle 45] (ST.south)++(0,0.15) |-node{Yes} (stop);	
\begin{pgfonlayer}{background}
\path (start.west |- start.north)+(-0.7,0.24) node (a) {};
\path (FR.east |- FR.south)+(6.0,-0.24) node (b) {};
\path[fill=green!10,rounded corners, draw=black!50, dashed](a) rectangle (b); 
\node [right of=start, node distance=4.1cm,yshift=0.5cm,xshift=0.9,align=center] (d2) {\large\bf PENALTY METHOD}; 
\end{pgfonlayer} 

\begin{pgfonlayer}{background}
\path (BFGS2.west |- BFGS2.north)+(-0.55,0.24) node (a) {};
\path (stop.east |- stop.south)+(6.25,-0.24) node (b) {};
\path[fill=green!10,rounded corners, draw=black!50, dashed](a) rectangle (b); 
\node [right of=BFGS2, node distance=6cm,yshift=0.5cm,xshift=-0.9cm,align=center] (d2) {\large \bf EXTENSION}; 
\end{pgfonlayer} 

\end{tikzpicture}}
	\caption{Flowchart of the robust penalty method.} 
	\label{fig:rpm}
\end{figure}

Fig.~\ref{fig:rpm} shows the flowchart of the proposed robust penalty method. The traditional penalty method is followed till the solution enters the feasible region, thereafter using the proposed extension of the method till convergence is achieved.

In MPC, optimization is performed at each sample time. For the first iteration of the closed-loop simulation, we take the initial guess for input $U_N =[0, 0,0, \dots, 0]$ where $N$ represents the prediction horizon, and the size of $U_N$ depends on the pre-defined prediction horizon ($N$). 
In subsequent iterations, the solver's initial guess will be the solution of the previous iteration. This will ensure a warm startup for the solver. If the initial point is in a feasible region, then the penalty is zero, and an unconstrained optimization problem is solved. 
Since the process is not under violation, the optimization problem is solved until the convergence criteria are fulfilled or any constraint is violated. If the former criteria terminates the problem, then the solution is in the feasible region. Else it lies on the boundary of the feasible region. 
In contrast to the above situation, if the initial guess is in the infeasible region, a penalty is added on the violated constraints, and Algorithm~\ref{algo:rpm} is followed until the point moves into the feasible space where 
the penalty becomes zero. 
Solving the problem even when the penalty is zero is an extension of the penalty method. This is independent of the initial guess, and the point will converge to a solution irrespective of whether the initial point was in a feasible or infeasible region. 
\section{Embedded Implementation}
\label{sec:embedded:lmpc}

\begin{figure}[h]
	\centering
	\includegraphics[scale=0.47]{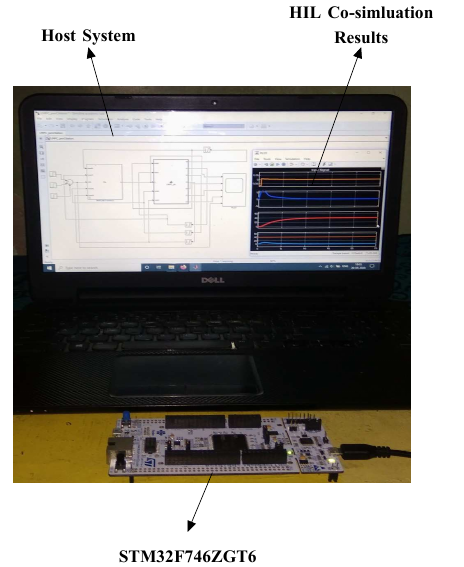}
	\caption{Hardware set-up for the MPC to perform HIL co-simulation using STM32 microcontroller.}
	\label{fig:HIL:set:up}
\end{figure}

This section describes the embedded implementation of linear MPC for citation aircraft model using ASM, IPM, and proposed robust penalty method as QP solvers.
The hardware used for implementation is STM Nucleo-144 development board with STM32F746ZGT6 MCU. It is an ARM 32-bit Cortex M7 microcontroller running at $216$~\unit{MHz}. It has $1024$~\unit{Kb} of program memory and $320$~\unit{Kb} of SRAM. Fig.~\ref{fig:HIL:set:up} shows the HIL co-simulation set-up. 
%
Fig.~\ref{fig:result:flow2} shows the design flow of the embedded implementation. The top left block shows the design steps carried in the \matlab~environment. The bottom left block shows the HIL co-simulation by interfacing with the Simulink environment. The blocks at right depict the function call of the respective QP solvers into the \matlab~environment for performance comparison. 

\begin{itemize}
	\item System modeling: We consider a linear discrete-time state-space model (as in~\eqref{eq:dlti}) of the citation aircraft system/plant under consideration to design MPC. Here, we consider a case study of the citation aircraft system. The citation aircraft system is inherently nonlinear. The model is linearized about equilibrium points and is transformed into a discrete-time model. This linearized model is used in the formulation of a linear MPC.
	\item MPC problem: the MPC problem is constructed (as in~\eqref{eq:dlti}) for the aircraft system by considering the reference tracking formulation (as in~\eqref{eq:cftoc}). Subsequently, the QP matrices/vectors, \ie, $H, f, G, w$ as in~\eqref{eq:qp} are generated in double-precision floating-point numbers.
	\item QP solver: the 
	algorithms are \cpp programs, in single-precision floating-point format i.e, float and synthesized using STM32CubeIDE,
	with GNU ARM compiler, which is used in \matlab ~by creating its MEX file. 
	\item Software-in-the-loop (SIL) testing/verification: from this step, we get an idea about the computational burden and memory demand. After getting the desired performance, we deployed the code on the hardware. 
	\item HIL co-simulation: we designed the simulator model in the \matlab/Simulink. For performing HIL co-simulation, Hardware development board STM32 Nucleo F746ZG is used for the implementation of the QP solver. 
	Processor-in-the-loop (PIL) communication interface is serial communication with the universal synchronous/asynchronous receiver/transmitter (USART2). 
	We used the Simulink embedded coder for deploying RPM solver and for ASM/IPM, we use the {\tt S function} block in the Simulink model. 
	We deployed the QP solver on the microcontroller, 
	We built a subsystem of the closed-loop model on the microcontroller, which generates the PIL block. This replaces the subsystem block. The closed-loop results obtained are discussed in section~\ref{sec:result}
\end{itemize}
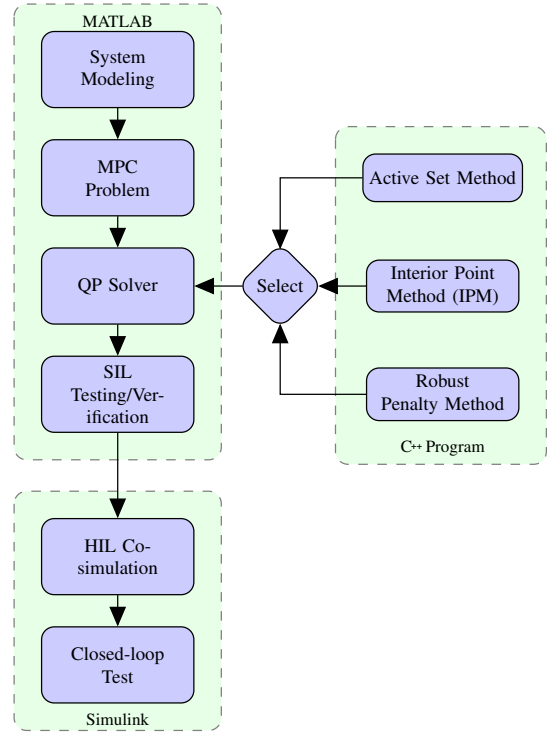
\begin{figure}[htbp]
	\centering
	\vspace*{-4mm}
	\scalebox{1.2}{\pgfdeclarelayer{background}
\pgfdeclarelayer{foreground}
\pgfsetlayers{background,main,foreground}
\tikzstyle{block} = [rectangle, draw, fill=blue!20,  text centered, rounded corners, minimum height=4em,minimum width=8em,text width=5em]
\tikzstyle{bck} = [rectangle, draw, fill=blue!20,  text centered, rounded corners, minimum height=2.5em,minimum width=8em]
\tikzstyle{dia} = [diamond, draw, fill=blue!20,  text centered, rounded corners, minimum height=2.5em,minimum width=4em]
\tikzstyle{line} = [draw, -latex']
\begin{tikzpicture}[node distance = 2.0cm, auto,scale=0.6, every node/.style={transform shape}]
\node [block,align=center] (model) {System Modeling};	
\node [block, below of=model,align=center] (mpc) {MPC Problem};	
\node [block, below of=mpc,align=center] (code) {QP Solver};
\node [block, below of=code,align=center] (SIL){SIL Testing/Verification};
\node [block, below of=SIL,align=center,node distance = 3.0cm] (hw) {HIL {Co-simulation}};
\node [block, below of=hw,align=center] (closed) {Closed-loop Test};
\node [dia, right of=code,align=center,node distance = 3cm] (select) {Select};
\node [bck, right of=select,align=center,node distance = 3cm] (mex) {Interior Point  \\ Method (IPM)};	
\node [bck, below of=mex,align=center,node distance = 2.0cm] (epm) {Robust \\ Penalty Method};		
\node [bck, above of=mex,align=center, node distance = 2.0cm] (quad) {Active Set Method};
\draw [-triangle 45] (model) -- (mpc);
\draw [-triangle 45] (mpc) -- (code);
\draw [-triangle 45] (code) -- (SIL);
\draw [-triangle 45] (SIL) -- (hw);	
\draw [-triangle 45] (hw) -- (closed);
\draw [-triangle 45] (select.west)++(0.1,0) -- (code);
\draw [-triangle 45] (mex) -- ++(-2.3,0);      
\draw [-triangle 45] (epm) -| (select.south)++(0,0.1);
\draw [-triangle 45 reversed] (quad) -| (select.north)++(0,0.25);	
\begin{pgfonlayer}{background}
\path (model.west |- model.north)+(-0.5,0.5) node (a) {};
\path (SIL.east |- SIL.south)+(0.5,-0.5) node (b) {};
\path[fill=green!10,rounded corners, draw=black!50, dashed](a) rectangle (b);    
\node [above of=model, node distance=0.9cm,yshift=0cm,align=center] (d2) {\small MATLAB};         
\end{pgfonlayer} 

\begin{pgfonlayer}{background}
\path (hw.west |- hw.north)+(-0.5,0.5) node (a) {};
\path (closed.east |- closed.south)+(0.5,-0.5) node (b) {};
\path[fill=green!10,rounded corners, draw=black!50, dashed](a) rectangle (b);    
\node [below of=closed, node distance=1cm,yshift=-0cm,align=center] (d2) {\small Simulink};         
\end{pgfonlayer} 

\begin{pgfonlayer}{background}
\path (quad.west |- quad.north)+(-0.5,0.5) node (a) {};
\path (epm.east |- epm.south)+(0.5,-0.8) node (b) {};
\path[fill=green!10,rounded corners, draw=black!50, dashed](a) rectangle (b);    
\node [below of=epm, node distance=1cm,yshift=-0cm,align=center] (d2) {\small \cpp Program};         
\end{pgfonlayer} 


\end{tikzpicture}}
	\caption{Design flow of embedded LMPC.} 
	\label{fig:result:flow2}
\end{figure}
\section{ Results} 
\label{sec:result}
This section presents the performance comparison of proposed method against ASM and IPM for benchmark QP problem and citation aircraft model. 
\subsection{Benchmark QP problem: qptest}
QP solver requires a standard QP problem for the performance evaluation. We used the qptest problem, which is a standard QP problem adopted from the 1999 Maros and Meszaros repository~\cite{maros:1999:repository}.
\begin{subequations}
\label{eq:qptest}
\begin{align}
\label{eq:qptest:obj}
&\text{min f(x,y)} = 4 + 1.5x - 2y + \nonumber\\& \hspace{2cm}\frac{1}{2}(8x^{2} + 2xy + 2 yx + 10y^{2}),\\
& \text{s.t.} \nonumber \\
\label{eq:qptest:const}
&2x + y \ge 2,\\
\label{eq:qptest:const2}
&-x + 2y \le 6,\\
\label{eq:qptest:const3}
&0 \le x \le 20,~y\ge 0,
\end{align}
\end{subequations}
The performance comparison of the solvers for the qptest is shown in Table~\ref{tab:table}. We see that the execution time for IPM and RPM is comparable. However, the ASM takes a longer time to reach the same optimal value. 
\begin{table}[h!]
\begin{center}
	\centering
	\setlength{\tabcolsep}{4.5pt}
	\renewcommand*{\arraystretch}{1.1}
	\caption{Performance comparison of three QP solvers for benchmark QP problem.}
	\label{tab:table}
	\begin{tabular}{l|c|c}
		\toprule 
		\textbf{Solver} & \textbf{Execution Time [s]} & \textbf{Optimal Value}\\
		\midrule 
		Active Set Method & 0.05&4.3718750 \\
		Interior Point Method& 0.004 & 4.3718750\\
		Robust Penalty Method & 0.004 &4.3718750 \\
		\bottomrule 
	\end{tabular}
\end{center}
\end{table}
%
\subsection{Citation Aircraft Control using MPC}
\label{subsec:aircraft}
To show the efficiency of the developed RPM QP solver, we present a case study of citation aircraft control problem~\cite[Chapter 3]{maciejowski:2002:predictive}. The longitudinal dynamics of a citation aircraft model are considered here using 
its linearized model with constant speed approximation. 
An aircraft is considered to be moving at an altitude of 5000~\unit{m} and a constant speed of 128.2~\unit{m/s}. Model has 4 states as $x =$ [angle of attack (\unit{$^{\circ}$}), pitch angle (\unit{$^{\circ}$}), altitude (\unit{m}), altitude rate (\unit{m/s})]. The control input is the elevator angle (\unit{$^{\circ}$}). The outputs of interest are: $y =$ [pitch angle (\unit{$^{\circ}$}), altitude (\unit{m}), altitude rate (\unit{m/s})]. We consider a discrete-time linear-time invariant state-space model as presented and parameter values in~\cite[Chapter 3]{maciejowski:2002:predictive}. 
\begin{subequations}
\label{eq:tank:dlti}
\begin{align}
x(t+T_s) = A x(t) + B u(t),\\
y(t) = Cx(t) + Du(t),
\end{align}
\end{subequations} 
where,
\begin{align}
\nonumber
&A= \begin{bmatrix}
0.240 & 0 & 0.1787 & 0\\
-0.372 & 1.000 & 0.270 & 0\\
-0.990 & 0 & 0.138 & 0\\
-48.935 & 64.100 & 2.399 & 1.000\\
\end{bmatrix},\;\;
B=\begin{bmatrix}-1.234\\
-1.438\\
-4.482\\
-1.799\\
\end{bmatrix}, 
\end{align}
\begin{align}
\nonumber
&C= \begin{bmatrix}
0 & 1.000 & 0 & 0\\
0 & 0 & 0 & 1.000\\
-128.200 & 128.200 & 0 & 0\\
\end{bmatrix},\;\;
D=\begin{bmatrix} 0\\0\\0 \end{bmatrix}.
\end{align}

The elevator angle has the constraints of $\pm 15^{\circ}$ and the elevator slew rate is limited to $\pm 30^{\circ}$~\unit{s}. These limitations are introduced due to the inherent design aspect of the aircraft. In addition to this, the passenger's comfort level is maintained by limiting the pitch angle between $\pm 20^{\circ}$.\\
\begin{figure*}[h!]
\centering
\includegraphics[scale=0.4]{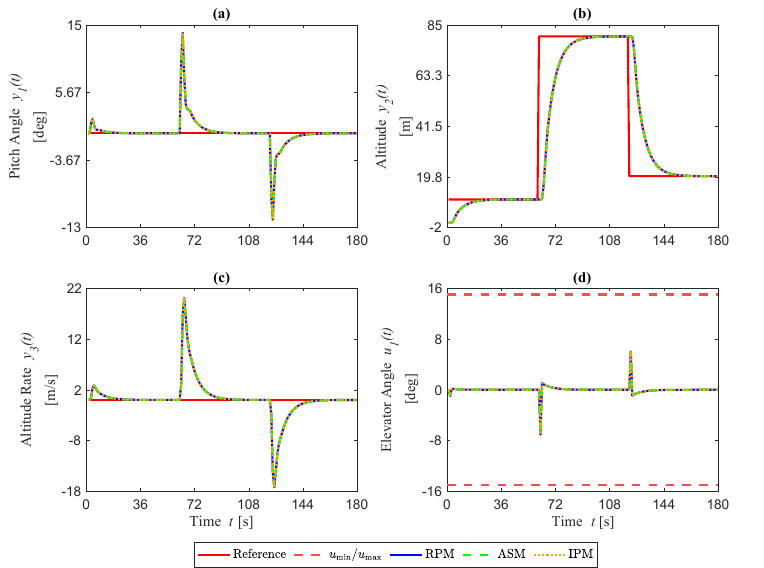}
\caption{Performance evaluations of three QP solvers used in MPC for citation aircraft control problem with $N = 10$.} 
\label{fig:result:comp}
\end{figure*}
To compare the performance of different methods, we plot the relative cumulative suboptimality (RCSO), relative to ASM. The RCSO is obtained as:
\begin{subequations}
	\label{eq:rcso}
	\begin{align}
	\label{eq:rcso:calc}
	\rm RCSO_{(\cdot),T_{\rm sim}} = \Biggl\lvert \frac{\rm DR_{(\cdot),T_{\rm sim}}\ -\ \rm DR_{\rm ASM,T_{\rm sim}}}{ \rm DR_{\rm ASM,T_{\rm sim}}} \Biggr\rvert,
	\end{align}
\end{subequations}
where $T_{\rm sim}$ is the number of time steps in the HIL-simulation.
Table~\ref{tab:table1} shows the RCSO comparison. 
A prediction horizon of 5 and 10 is set for this application. Moreover, the RPM is also tested for prediction horizon of 20 and 30, which resulted in a minimal increase of computational time and with no significant improvement of response.
The result is generated for the convergence tolerance of 1E-6 for all solvers. However, it can be changed as per the application and respective requirements. 
\begin{table}[htbp]
	\begin{center}
		\caption{Relative suboptimality of the different examples. Note: small value is better}
		\label{tab:table1}
		\setlength{\tabcolsep}{4.0pt}
		\renewcommand*{\arraystretch}{1}
		\begin{tabular}{c|c|c}
			\toprule 
			\textbf{QP} & \textbf{Prediction} & \textbf{RCSO}\\
			\textbf{Solver}& \textbf{Horizon} & \\
			\midrule 
			Active Set & 5 & 00e+00\\
			Method & 10 & 00e+00\\\midrule
			Interior Point & 5 & 2.31e-02\\
			Method & 10 & 5.12e-03\\\midrule		
			Robust & 5 & 4.73e-03\\
			Penalty Method & 10 & 4.12e-05\\
			\bottomrule 
		\end{tabular}
	\end{center}
\end{table}

We observed that the RCSO of RPM is less compared to IPM. 
Table~\ref{tab:table2} demonstrates the performance of solver with a different predictive horizon. We see that RPM is highly competent and outperforms IPM and ASM. Since the execution of RPM will depend on the slack tolerance, smaller the slack value, the higher will be the error and vice versa. 

\begin{table}[htbp]
\begin{center}
	\caption{Average execution time (in seconds) on STM Nucleo-144 board for three QP solvers with different prediction horizons.}
	\label{tab:table2}
	\setlength{\tabcolsep}{4.5pt}
	\renewcommand*{\arraystretch}{1}
	\begin{tabular}{c|c|c|c}
		\toprule 
		\textbf{Prediction} & \textbf{Active Set} & \textbf{Interior Point}& \textbf{Robust Penalty} \\
		\textbf{Horizon} & \textbf{Method} & \textbf{Method}& \textbf{Method}\\
		\midrule 
		\textbf{5} & 0.0071 & 0.0051& 0.0046 \\
		\textbf{10} & 0.0075 & 0.0051& 0.0046\\
		\textbf{20} & 0.0075 & 0.0051& 0.0054\\
		\textbf{30} & 0.0085 & 0.0056& 0.0054\\
		\textbf{50} & 0.016 & 0.0057&0.0055 \\
		\bottomrule 
	\end{tabular}
\end{center}
\end{table}

The variation of computational time for different slack tolerance is shown in Table~\ref{tab:table3}. We see that the increase in execution time is not very significant. Moreover, slack tolerance can be a user-defined parameter, depending on the application.
\begin{table}[htbp]
\begin{center}
	\caption{Average execution time (in seconds) on STM Nucleo-144 board for robust penalty method solver with different slack tolerances when $N = 10$.} 
	\label{tab:table3}
	\begin{tabular}{c|c}
		\toprule 
		\textbf{Slack Tolerance} & \textbf{Execution Time [s]} \\
		\midrule 
		1E-4 & 0.0019 \\
		1E-6 & 0.0021 \\
		1E-9 & 0.0024 \\
		1E-12 & 0.0028 \\
		1E-15 & 0.0028 \\
		1E-18 & 0.0029 \\
		\bottomrule 
	\end{tabular}
\end{center}
\end{table}
In Table~\ref{tab:tablecon}, we show the comparison of convergence tolerance for the three solvers. RPM performs better than IPM and ASM for the lower convergence tolerance, but the execution time of RPM increases more significantly than the other two. Since the performance of MPC doesn't depend on the extent to which QP is solved~\cite{wang:2009:fast}, this does not have major contribution towards the final performance. 
\begin{table}[htbp]
\begin{center}
	\caption{Average execution time (in seconds) on STM Nucleo-144 development board for three QP solvers used in MPC with different optimality tolerances.}
	\label{tab:tablecon}
	\setlength{\tabcolsep}{4.5pt}
	\renewcommand*{\arraystretch}{1}
	\begin{tabular}{c|c|c|c}
		\toprule 
		\textbf{Optimality} & \textbf{Active Set} & \textbf{Interior Point}& \textbf{Robust Penalty} \\
		\textbf{Tolerance} & \textbf{Method} & \textbf{Method}& \textbf{Method}\\
		\midrule 
		\textbf{1E-4} & 0.0105 & 0.0116 & 0.0041 \\
		\textbf{1E-6} & 0.0106 & 0.0121 & 0.0046\\
		\textbf{1E-9} & 0.0108 & 0.0126 & 0.0011 \\
		\textbf{1E-12} & 0.0111 & 0.0133 & 0.0028 \\
		\bottomrule 
	\end{tabular}
\end{center}
\end{table}

Fig.~\ref{fig:result:comp} displays the 
results for reference tracking using the three solvers. It is observed that all the three methods track the reference accurately while satisfying the constraints. 

\section{Conclusion}
\label{sec:conclusion}
This paper has presented a robust penalty method approach for solving the linear MPC. 
Our extension makes the penalty method robust to any initial guess, unlike the traditional penalty method. 
The proposed QP solver is implemented on a low-cost STM32 microcontroller and its performance is compared with that of state-of-the-art QP solvers such as the ASM and IPM. 
The HIL co-simulation results of the three QP solvers are presented for control of aircraft problem. A detailed analysis of computational complexity in terms of hardware execution time and steady-state error is presented. The results show that the proposed RPM outperforms the ASM and IPM solvers in execution time. We suggest some directions that can be explored for further research:
\begin{itemize}
\item The proposed QP solver needs to be tested for complex and large scale QP problem for further analysis.
\item This solver can be extended to solve nonlinear MPC due to its capability to handle nonlinear constraints.
\end{itemize}

\section*{Acknowledgment}
We gratefully acknowledge the support from the R\&D center of Embedded lab in the Instrumentation and Control department at COEP. Vaishali Patne acknowledges the contribution of the Department of Science and Technology, Govt. of India, under Women Scientist Scheme-A (WOS-A/ET-$120$/$2018$). Deepak Ingole would like to thank the financial support of the Moonshot-FLEX project of the Flemish Government.
\bibliographystyle{IEEEtran}
\Urlmuskip=0mu plus 1mu\relax
\bibliography{root} 
\end{document}